\documentclass[aps,preprint,pra]{revtex4-1}
\usepackage{graphicx}
\usepackage{amssymb}
\usepackage{amsmath}
\usepackage{color}
\usepackage{enumerate}
\usepackage{bm}
\usepackage[utf8]{inputenc}
\begin{document}
\title{ Controlling Dipolar Interaction Effect in Two-Dimensional Magnetic Nanostructures}
\author{Manish Anand}
\email{itsanand121@gmail.com}
\affiliation{Department of Physics, Bihar National College, Patna University, Patna-800004, India.}

\date{\today}

\begin{abstract}
 We investigate the dependence of magnetic properties on the out-of-plane disorder strength $\Delta$, dipolar interaction strength $h^{}_d$ in two-dimensional  ($l^{}_x\times l^{}_y$) ensembles of nanoparticles using numerical simulations. Such positional defects are redundantly observed in experiments. The superparamagnetic character is dominant with negligible and weak interaction strength $h^{}_d$, irrespective of $\Delta$ and aspect ratio of the system $A^{}_r=l^{}_y/l^{}_x$. The double-loop hysteresis curve,  characteristics of antiferromagnetic coupling dominance, emerges with large $h^{}_d$ and $\Delta(\%)\leq5$ in the square-like nanoparticles' assays. Remarkably, the dipolar interaction of sufficient strength drives the magnetic order from antiferromagnetic to ferromagnetic with large $\Delta$ and $A^{}_r\leq4.0$, resulting in an enhancement in the hysteresis loop area. On the other hand, the ferromagnetic coupling gets increased with $h^{}_d$ in systems with huge $A^{}_r$. Consequently, the hysteresis loop is enormous, even with moderate $h^{}_d$. The variation of the coercive field $\mu^{}_oH^{}_c$, remanence $M^{}_r$, and amount of heat released $E^{}_H$ (due to the hysteresis) with these parameters also suggests the transformation of nature dipolar interaction. They are significant even with large $h^{}_d$ and smaller $A^{}_r$, indicating the antiferromagnetic coupling dominance. Interestingly, there is an enhancement in these with $\Delta$ and large $h^{}_d$ due to ferromagnetic interaction. Notably, they are very significant even with moderate $h^{}_d$ in the highly anisotropic system and external field along the long axis of the sample. These results could help the experimentalist in explaining the unusual hysteresis characteristics observed in such systems and should also be beneficial in diverse applications such as data storage, magnetic hyperthermia, etc.

\end{abstract}
\maketitle
\section{Introduction}
Two-dimensional ensembles of magnetic nanoparticles (MNPs) are one of the essential classes of nanostructures due to their exciting physics and various technological applications~\cite{tan2017recent,farhan2013,puntes2004,leo2018,ge2020,wang2019,puntes2004}. Their importance for fundamental research stems from multiple factors such as low dimensionality and a significant contribution from increased surface atoms, resulting in intriguing magnetic properties not observed in the bulk systems~\cite{jordanovic2015,verba2020,yakovlev2017,micheletto1995}. Consequently, they have tremendous prospects in diverse applications such as spintronics, ultra-high-density magnetic recording media, magnetic hyperthermia, etc.~\cite{valdes2021,sarella2014,anand2018}.

The dipolar interaction plays an essential role in stabilizing magnetic order and determining the characteristics of such a system~\cite{hauschild1998,vega2019,weis2002}. The dipolar interaction is long-ranged and anisotropic; it has diverse effects on various properties of vital importance such as hysteresis, collective behaviour, blocking temperature, relaxation, etc.~\cite{allia1999,masunaga2016,anand2021magnetic}. Geng {\it et al.} systematically investigated the variation of remanence as a function of various parameters such as particle size, interparticle interaction, anisotropy in the ensemble of cobalt ferrite MNPs~\cite{geng2016}. They observed reduction in the remanence due to dipolar interaction. 
Poddar {\it et al.} synthesized monolayers and multilayers of magnetite MNPs and performed the ac and dc magnetic susceptibility measurements~\cite{poddar2002}. The dipolar interaction induces spin-glass-like slowing down of magnetic relaxation in such cases. Bahiana {\it et al.} investigated the dipolar interaction effect on magnetic ordering in a two-dimensional array of MNPs using Metropolis Monte Carlo (MMC) technique~\cite{bahiana2004}. The dipolar interaction is found to induce antiferromagnetic coupling at low temperatures. In recent work, we investigated the hysteresis properties in ordered arrays of MNPs as a function of dipolar interaction strength, aspect ratio, external magnetic field directions using the kinetic Monte Carlo (kMC) algorithm~\cite{anand2021jmmm}. Interestingly, the dipolar interaction of sufficient strength induces antiferromagnetic in the square sample while ferromagnetic coupling in a highly anisotropic system. The dipolar interaction also strongly affects the ground state morphology in such a system. For example, the minimum energy state is antiferromagnetic in a square array of MNPs~\cite{macisaac1996,de1997}. In contrast, the dipolar interaction promotes the ferromagnetic arrangement of spins in a triangular lattice~\cite{politi2006}. In a simple cubic lattice, the minimum energy spins' configuration is dictated by antiferromagnetic coupling~\cite{luttinger1946}. While in face centered cubic arrangement of MNPs, the ground state is ferromagnetic~\cite{luttinger1946}.   

Moreover, the dipolar interaction also significantly affects the hysteresis loop area, one of the essential quantifiers for magnetic hyperthermia application~\cite{russier2001,xue2006,haase2012,takagaki2005,tan2010}. V. Russier {\it et al.} investigated the hysteresis in the two-dimensional array of MNPs~\cite{russier2001}. The dipolar interaction is found to have a detrimental effect on the hysteresis loop area. Xue {\it et al.} studied the hysteresis characteristics in an infinite two-dimensional hexagonal array of MNPs~\cite{xue2006}. The hysteresis loop area gets decreased because of dipolar interaction. Haase and Nowak also found a detrimental effect of dipolar interaction on the amount of heat dissipated in the hysteresis process~\cite{haase2012}. Takagaki {\it et al.} also observed a strong effect of the dipolar interaction on the hysteresis in the square arrangement of nanomagnet~\cite{takagaki2005}. Tan {\it et al.}  also observed a decrease in hysteresis loop area and tunnel magnetoresistance in the two-dimensional MNPs arrays at temperature $T=0$ K due to the dipolar interaction~\cite{tan2010}. 
In recent work, we also observed a strong dipolar interaction effect on the hysteresis loop area in ordered arrays of MNPs~\cite{anand2020}. The dipolar interaction also affects the magnetotransport properties in MNPs ensembles, an essential measure for spintronics based applications~\cite{bupathy2019,kechrakos2008,dugay2011}.

The above discussion clearly shows that the dipolar interaction has varied effects on various systematic properties in MNPs arrays. In addition, the dipolar interaction may promote ferromagnetic or antiferromagnetic coupling between the magnetic moments depending on the interparticle separation and the geometrical arrangement of MNPs. 
Experimentally fabricated two-dimensional arrays of magnetic nanoparticles are generally not free from defects~\cite{gallina2020,riedinger2017,jensen2003,wiedwald2004}. One of the prominent classes of defects observed in such systems is out-of-plane disorder~\cite{wiedwald2004}. In such a case, all the MNPs in the arrays do not lie in a $xy$-plane (say); a few of them are also scattered along the $z$-axis.
Therefore, it is essential to understand the effect of such disorder on magnetic hysteresis and provide a theoretical basis to explain so-observed unexpected magnetic behaviour. These studies are of immense importance for efficient use of such systems in spintronics-based applications, digital data storage, and other related applications where such nanostructures are ubiquitous. Thus motivated, in the present work, we systematically investigate the effect of out-of-plane disorder strength, dipolar interaction, system size, aspect ratio and external magnetic field directions on the hysteresis mechanism in the two-dimensional MNPs ensembles using kinetic Monte Carlo simulation.

\section{Theoretical Framework}
\begin{figure}[!htb]
	\centering\includegraphics[scale=0.40]{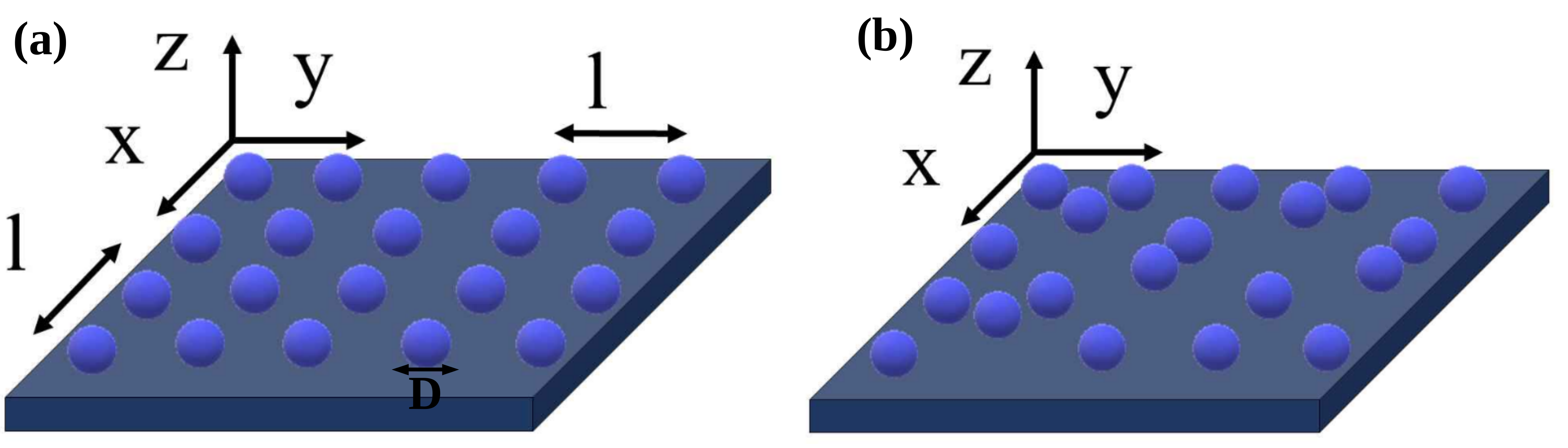}
	\caption{(a) Schematic of two-dimensional perfectly ordered arrays of magnetic nanoparticles. (b) Schematic of arrays of nanoparticles with out-of-plane positional disorder.}
	\label{figure1}
\end{figure}

We consider two-dimensional arrays ($l^{}_x\times l^{}_y$) of spherical and monodisperse MNPs. The particles are assumed to lie in $xy$-plane with a fraction of them scattered along the $z$-axis, as shown in schematic Fig.~(\ref{figure1}). Such positional defects are termed as out-of-plane disorder, advertently observed in experiments. Let the nanoparticle has diameter $D$ and the lattice spacing $l$ [see  Fig.~(\ref{figure1})]. 
The anisotropy energy $E^{}_K$ associated with such particle is given by~\cite{anand2016,carrey2011}
\begin{equation}
E^{}_K=K^{}_{\mathrm{eff}}V\sin^2\Phi
\end{equation} 
Here $K^{}_{\mathrm {eff}}$ is the anisotropy constant corresponding to uniaxial anisotropy vector $\vec{K}= K^{}_{\mathrm {eff}} \hat{k}$, $\hat{k}$ is the unit vector along with the anisotropy or easy axis. The nanoparticle has volume $V=\pi D^3/6$ and $\Phi$ is the angle between $\hat{k}$ and magnetic moment vector.

In an assembly, MNPs interact due to the dipolar interaction. The corresponding interaction energy $E^{}_{\mathrm {dip}}$ can be calculated using the following expression~\cite{anand2021thermal}
\begin{equation}
\label{dipole}
E^{}_{\mathrm {dip}}=\frac{\mu^{}_o\mu^2}{4\pi l^3}\sum_{j,\ j\neq i}\left[ \frac{\hat{\mu_{i}}\cdot\hat{\mu_{j}}-3\left(\hat{\mu_{i}}\cdot\hat{r}_{ij}\right)\left(\hat{\mu_{j}}\cdot\hat{r}_{ij}\right)}{(r^{}_{ij}/l)^3}\right].
\end{equation}
Where the permeability of free space is $\mu_{o}$; $\mu=M^{}_sV$ is the magnetic moment of a nanoparticle, $M^{}_s$ is the saturation magnetization. The $i^{th}$ and $j^{th}$ magnetic moments have unit vectors $\hat{\mu}_{i}$ and $\hat{\mu}_{j}$, respectively, separated by a distance $r^{}_{ij}$. $\hat{r}^{}_{ij}$ is the unit vector corresponding to $\vec{r}_{ij}$.

 
The corresponding field to the dipolar interaction is expressed as~\cite{anand2021thermal,tan2014}
\begin{equation}
\mu^{}_{o}\vec{H}^{}_{\mathrm {dip}}=\frac{\mu_{o}\mu}{4\pi l^3}\sum_{j,j\neq i}\frac{3(\hat{\mu}^{}_j \cdot \hat{r}_{ij})\hat{r}^{}_{ij}-\hat{\mu^{}_j} }{(r_{ij}/l)^3}.
\label{dipolar1}
\end{equation}
We model the dipolar interaction strength by a control parameter $h^{}_d=D^{3}/l^3$~\cite{tan2010}. It correctly captures the physics of dipolar interaction variation as the latter varies as $1/r^{3}_{ij}$, evident from Eq.~(\ref{dipole}) and Eq.~(\ref{dipolar1}). Therefore, the magnetically non-interacting system can be modelled with $h^{}_d=0$ while $h^{}_d= 1.0$ corresponds to the strongest dipolar interacting case.



The alternating magnetic field is applied to the system to study the magnetic hysteresis. It 
is given by~\cite{anand2020}
\begin{equation}
\mu^{}_{o}H=\mu^{}_oH^{}_{\mathrm {max}}\cos\omega t,
\label{magnetic}
\end{equation} 
where $\mu^{}_{o}H_{\mathrm {max}}$ is the amplitude of the external magnetic field; $\omega=2\pi\nu$ is the angular frequency. $\nu$ is the linear frequency, and $t$ is time. 
We can therefore write the expression for the total energy $E$ as  ~\cite{tan2014,anand2019}
\begin{equation}
E=K^{}_{\mathrm {eff}}V\sum_{i}\sin^2 \Phi^{}_i+\frac{\mu^{}_o\mu^2}{4\pi l^3}\sum_{j,\ j\neq i}\left[ \frac{\hat{\mu_{i}}\cdot\hat{\mu_{j}}-{3\left(\hat{\mu_{i}}\cdot\hat{r}_{ij}\right)\left(\hat{\mu_{j}}\cdot\hat{r}_{ij}\right)}}{(r_{ij}/l)^3}\right]-\mu^{}_o\mu\sum_{i}\hat{\mu}^{}_i\cdot\vec{H}
\end{equation}
Here $\Phi^{}_i$ is the angle between the anisotropy axis and the $i^{th}$ magnetic moment of the system. 

We use state of the art kinetic Monte Carlo simulation to investigate the magnetic hysteresis in the two-dimensional assembly of MNPs in the presence of the out-of-plane disorder.  To quantify the out-of-plane-disorder, we define a disorder strength parameter $\Delta(\%)=n^{}_z/N$, where $N$ is the total number of nanoparticles in the system and $n^{}_z$: the number of particles scattered along the $z$-axis (normal to the plane of the system). In the present work, we systematically analyzed the hysteresis response as a function of $\Delta (\%)$, dipolar interaction strength $h^{}_d$, aspect ratio $A^{}_r=l_y/l_x$, and the external magnetic field's direction. The kMC algorithm is a more suitable choice than the MMC technique to investigate the time-dependent properties such as dynamic hysteresis, relaxation characteristics, etc. It is because the dynamics are well captured in the kMC methods~\cite{chantrell2000}. Tan {\it et al.} implemented the same kMC procedures to investigate the spatial distribution of heat dissipation in the spherical assembly of MNPs~\cite{tan2014}. We have also used it to study various properties in arrays of dipolar interacting nanoparticles~\cite{anand2019,anand2020,anand2021thermal,anand2021tailoring}.

The area under the hysteresis curve equals the amount of heat disspated $E^{}_{H}$ during one complete cycle of external field variation, which can be evaluated using the following formula~\cite{anand2016}

\begin{equation}
E^{}_{H}=\oint M(H)dH,
\label{local_heat}
\end{equation}
The above intergral is evaluated over the entire period of the external magnetic field change. $M^{}(H)$ is the magnetization of system at the applied magnetic field $H$. 
\section{Simulations Results}
The nanoparticles are assembled in the two-dimensional  ($l^{}_x\times l^{}_y$) lattice [$xy$-plane] with fraction of them scattered along $z$-direction. We have considered following values of simulations parameters: $D=8$ nm, $K^{}_{\mathrm {eff}}=13\times10^{3}$ $\mathrm {Jm^{-3}}$, $M^{}_s=4.77\times10^{5}$ $\mathrm {Am^{-1}}$, $N=400$, $\mu^{}_oH^{}_{\mathrm {max}}=0.10$ T, $\nu=10^5$ Hz, and $T=300$ K. Five typical values of system sizes viz. $l^{}_x\times l^{}_y=20\times20$, $10\times40$, $4\times100$, $2\times200$, and  $1\times400$ are considered. These correspond to $A^{}_r(=l^{}_y/l^{}_x)=1.0$, 4.0, 25.0, 100.0, 400.0, respectively. We have varied $\Delta (\%)$ from 5 to 50 to extensively investigate the effect of out-of-plane positional disorder on the hysteresis characteristics. The dipolar interaction strength $h^{}_d$ is varied from 0 to 1.0. We have also assumed anisotropy axes to be randomly oriented in three-dimensional space to mimic the real system. The oscillating magnetic field is applied along $x$ and $y$-direction.

\begin{figure}[!htb]
	\centering\includegraphics[scale=0.50]{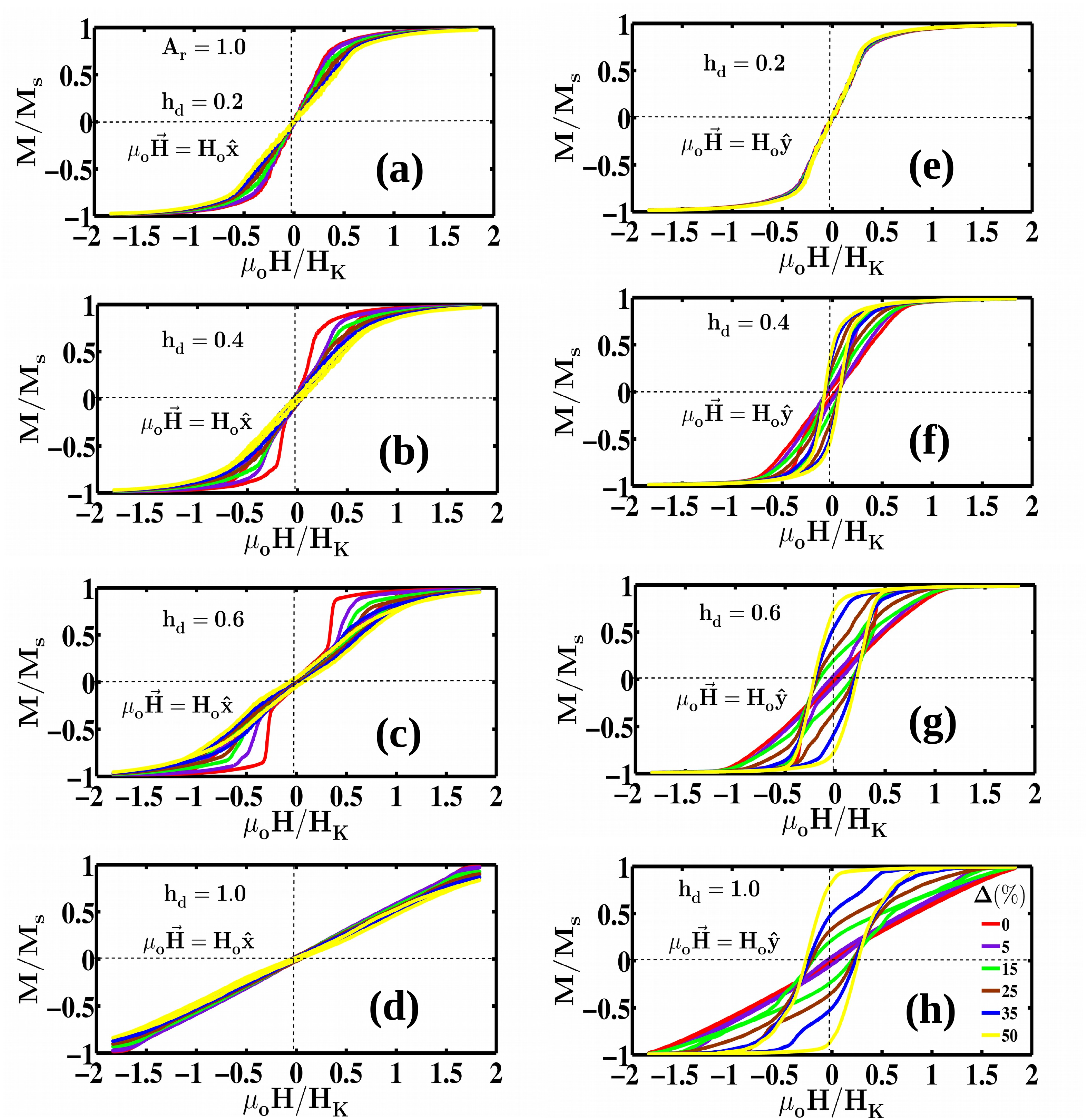}
	\caption{Magnetic hysteresis curves as a function of out-of-plane disorder strength $\Delta$ in square arrays ($A^{}_r=1.0$) of MNPs. We have considered four representative values of dipolar interaction strength $h^{}_d=0.2$ [(a), (e)], 0.4 [(b), (f)], 0.6 [(c), (g)] and 1.0 [(d), (h)]. The alternating magnetic field is applied along $x$ and $y$-directions. The dipolar of enough strength drives the system from antiferromagnetic to ferromagnetic state with significant $\Delta$, resulting an enhancement coercive field and remanence with $h^{}_d$. }
	\label{figure2}
\end{figure}

We first investigate the out-of-plane disorder effect on the hysteresis in square arrays of MNPs as a function of dipolar interaction strength. In Fig.~(\ref{figure2}), we plot the magnetic hysteresis curve as a function of $\Delta (\%)$, $h^{}_d$, and external magnetic field's direction with $A^{}_r=1.0$ ($l^{}_x\times l^{}_y=20\times20$). The alternating magnetic field is applied along $x$ and $y$-direction. The hysteresis curve without the disorder ($\Delta(\%)=0$) is also plotted for comparison. In all the hysteresis curves shown in the present work, the external magnetic field ($x$-axis) and the magnetization ($y$-axis) is rescaled by the single-particle anisotropy field $H^{}_K=2K^{}_{\mathrm {eff}}/M^{}_s$~\cite{carrey2011} and $M^{}_s$, respectively. The superparamagnetic character is dominant with weak and negligible dipolar interaction strength $h^{}_d\leq0.2$, irrespective of $\Delta$ and applied field's directions. Consequently, a characteristic hysteresis curve for superparamagnetic particles (S-shaped hysteresis curve with the  negligible value of coercive field and remanence) is observed in such a case. Nowak {\it et al.} also reported similar results using MMC simulations~\cite{nowak1994}. Interestingly, the double-loop hysteresis emerges with appreciable interaction strength $h^{}_d=0.4$ and $0.6$ in the absence of disorder ($\Delta(\%)=0$), reminiscent of antiferromagnetic coupling dominance. It can be explained by carefully analyzing the nature of the dipolar and anisotropy field as follows. The dipolar interaction forces the magnetic moments to align in the plane of the sample by inducing a biasing field in the planar arrangement of MNPs. In contrast, the  anisotropy field instigates the magnetic moments to orient in three-dimensional space as the anisotropy axes are considered to be randomly oriented in the present article. As a result, there is a linear variation of magnetization with the external field, provided it is weaker than the coercive field, i.e. $\mu^{}_o H/H^{}_K< 0.5$~\cite{carrey2011}. The dipolar field dictates the hysteresis behaviour for $\mu^{}_o H/H^{}_K>0.5$, resulting in a spontaneous orientation transition of magnetization. These results are in perfect agreement with the work of Yang {\it et al.}~\cite{yang2002}. It is also clearly evident that there is a strong effect of disorder on hysteresis response. Remarkably, the out-of-plane disorder puts the system from an antiferromagnetic to a ferromagnetic magnetic state with the field applied along the $y$-direction. Consequently, the hysteresis curve starts to open up near the small external magnetic field (coercive field), indicating enhancement of ferromagnetic coupling in the system. Therefore,  the area under the hysteresis curve increases with an increase in disorder strength $\Delta$.

\begin{figure}[!htb]
\centering\includegraphics[scale=0.50]{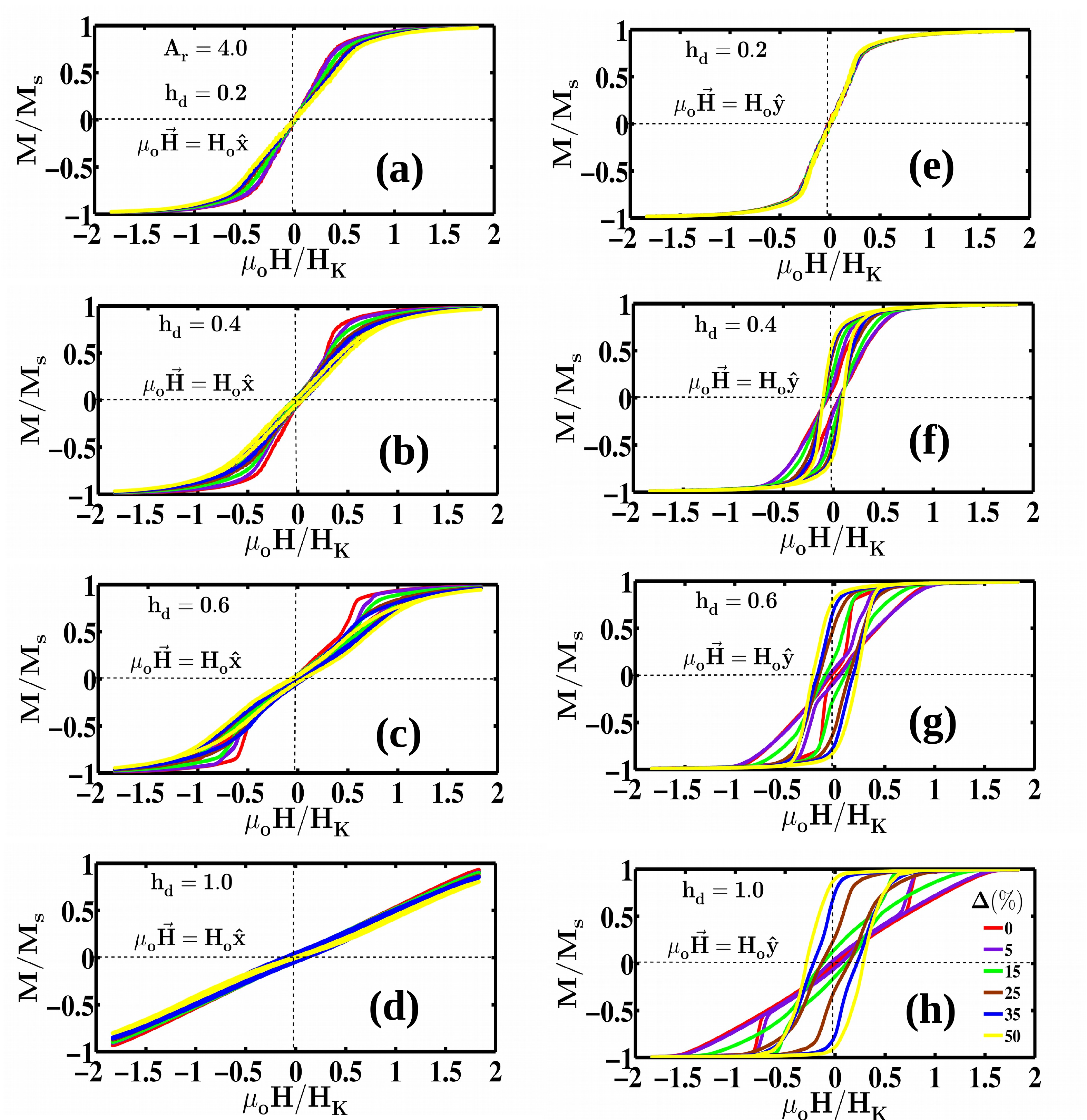}
\caption{Magnetic hysteresis in rectangular array of MNPs ($A^{}_r=4.0$) with disorder strength $\Delta$. We have also considered four typical values of interaction strength $h^{}_d=0.2$, 0.4, 0.6, 1.0. The external magnetic field is applied along $x$ [(a)-(d)] and y-directions [(e)-(h)]. The superparamagnetic character is dominant with $h^{}_d\leq0.2$. The double-loop hysteresis emerges for $h^{}_d\geq0.4$ and $\Delta(\%)\leq5$, indicating the antiferromagnetic coupling dominance. The nature of dipolar interaction changes from antiferromagnetic to ferromagnetic with $\Delta$.}
\label{figure3}
\end{figure}

Next, we analyze the magnetic hysteresis as a function of dipolar interaction and disorder strength in rectangular arrays of MNPs. Fig.~(\ref{figure3}) shows the hysteresis curve as a function of $\Delta$, $h^{}_d$ and applied field directions with $A^{}_r=4.0$ ($l^{}_x\times l^{}_y=10\times40$). All other parameters are the same as that of Fig.~(\ref{figure2}). In the presence of small $h^{}_d=0.2$, the superparamagnetic character is dominant in this case also. Notably, the non-hysteresis is observed with the field applied along the shorter axis of the sample ($x$-axis) for appreciable dipolar interaction strength ($h^{}_d>0.2$) and significant $\Delta$. The dipolar interaction promotes ferromagnetic coupling  with $\Delta$ and $h^{}_d$ along the long axis of the system ($y$-axis) because of its anisotropic nature. Consequently, the magnetization ceases to follow the external field when applied along the shorter axis ($x$-axis), resulting in a minimal hysteresis loop area. Interesting physics unfolds when the field is applied along the long axis of the system. The double-loop hysteresis emerges in an ordered ensemble of MNPs (negligibly small $\Delta$) with appreciable $h^{}_d$, characteristics of antiferromagnetic coupling. This observation is in perfect agreement with the work of Chen {\it et al.}~\cite{chen2017}. Remarkably, the system's rapid transition from antiferromagnetic to the ferromagnetic regime with $\Delta$, provided $h^{}_d$ is significant. Because the dipolar interaction increases the ferromagnetic coupling with an enhancement in the disorder strength, resulting in an enhanced hysteresis loop area. These observations are also in perfect agreement with the work of Cheng {\it et al.}~\cite{cheng2004magnetic}. These results clearly suggest that we can tune the nature of dipolar interaction from ferromagnetic to antiferromagnetic or vice versa by manoeuvering the disorder strength $\Delta$ and $h^{}_d$. These observations could be beneficial in manipulating the hysteresis loop area and other related properties of interest, which could be helpful in various applications such as magnetic hyperthermia and spintronics.

\begin{figure}[!htb]
\centering\includegraphics[scale=0.50]{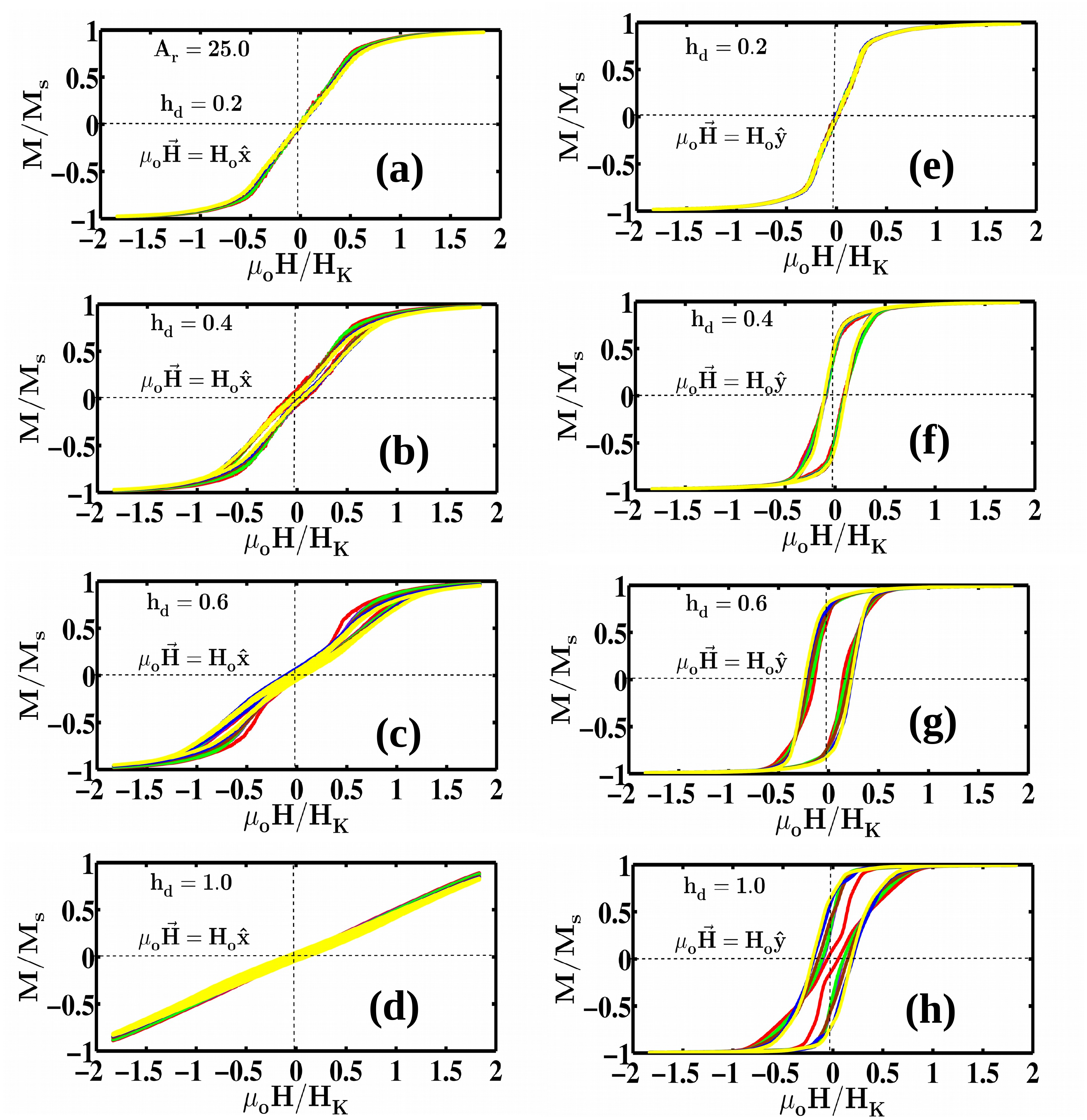}
\caption{The study of hysteresis as a function of dipolar interaction and disorder strength $\Delta$ in system with large aspect ratio $A^{}_r=25.0$. We have applied magnetic field along x [(a)-(d)] and y-axes [(e)-(h)]. Non-hysteresis is observed with the field applied along the $x$-direction. The dipolar interaction of enough strength promotes ferromagnetic coupling, resulting in an ehnanced hysteresis loop area when the field is applied along $y$-direction.}
\label{figure4}
\end{figure}

\begin{figure}[!htb]
\centering\includegraphics[scale=0.50]{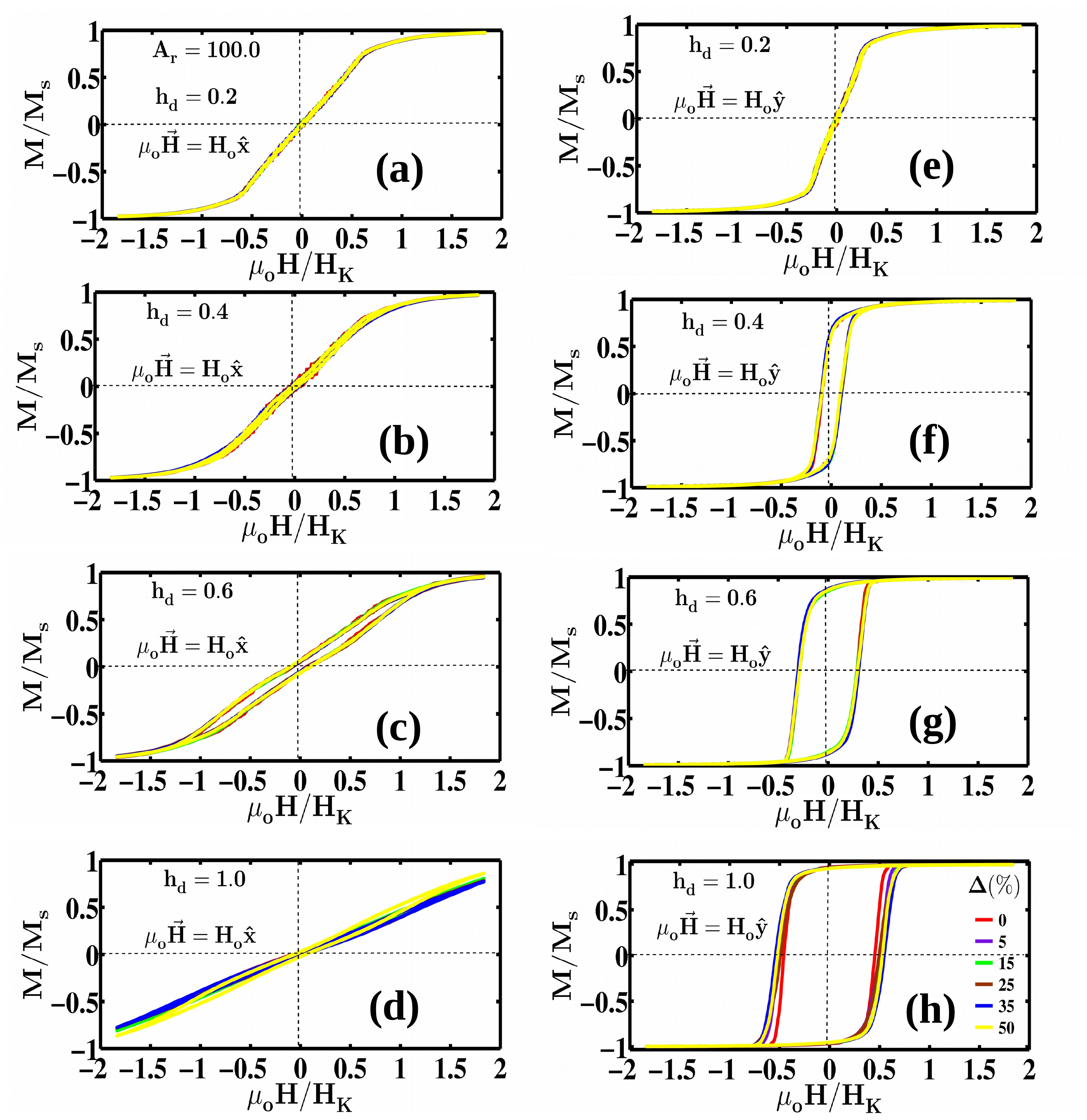}
\caption{The out-of-plane disorder and dipolar interaction dependence of the hysteresis in a very large anisotroic system ($A^{}_r=100.0$). There is a dominance of superparamagnetic behaviour for small $h^{}_d$ even with large $A^{}_r$. The magnetization ceases to follow the external field, resulting in non-hysteresis behaviour with the field along $x$-direction. In contrast, the hysteresis loop area is large even with moderate $h^{}_d$ when the external magnetic field is along the $y$-direction.}
\label{figure5}
\end{figure}

After that, we investigate the hysteresis mechanism in systems with large aspect ratios. In Fig.~(\ref{figure4}) and Fig.~(\ref{figure5}), we plot the hysteresis curve with $A^{}_r=25.0$ and $A^{}_r=100.0$, respectively. All the parameters are the same as that of Fig.~(\ref{figure3}). There is a negligible value of coercive field and remanence for weakly interacting nanoparticles, indicating superparamagnetic behaviour, irrespective of Ar and external field directions. The magnetization ceases to follow the external field ($\mu^{}_o\vec{H}=H^{}_o\hat{x}$) for appreciable $h^{}_d$, resulting in non-hysteretic behaviour. Notably, the dipolar interaction induces shape anisotropy or ferromagnetic coupling along the long axis with significant $A^{}_r$. Consequently, remanence and coercive field are significant even in the perfectly ordered system ($\Delta(\%)=0$) with the external magnetic field along the long axis of the sample, i.e. $y$-axis. Moreover, the hysteresis loop area is also significant and increases with $h^{}_d$. There is a weak dependence of hysteresis on the out-of-plane disorder as anticipated. The ferromagnetic coupling strength has already at its maximum even with $\Delta(\%)=0$ as the system is highly anisotropic. These observations clearly indicate that dipolar interaction of sufficient strength puts the underlying system from superparamagnetic to the ferromagnetic regime even at $T=300$ K and $\Delta(\%)$=0 (ordered arrays). Therefore, there is a huge hysteresis loop area with the external field along the $y$-direction.

\begin{figure}[!htb]
\centering\includegraphics[scale=0.50]{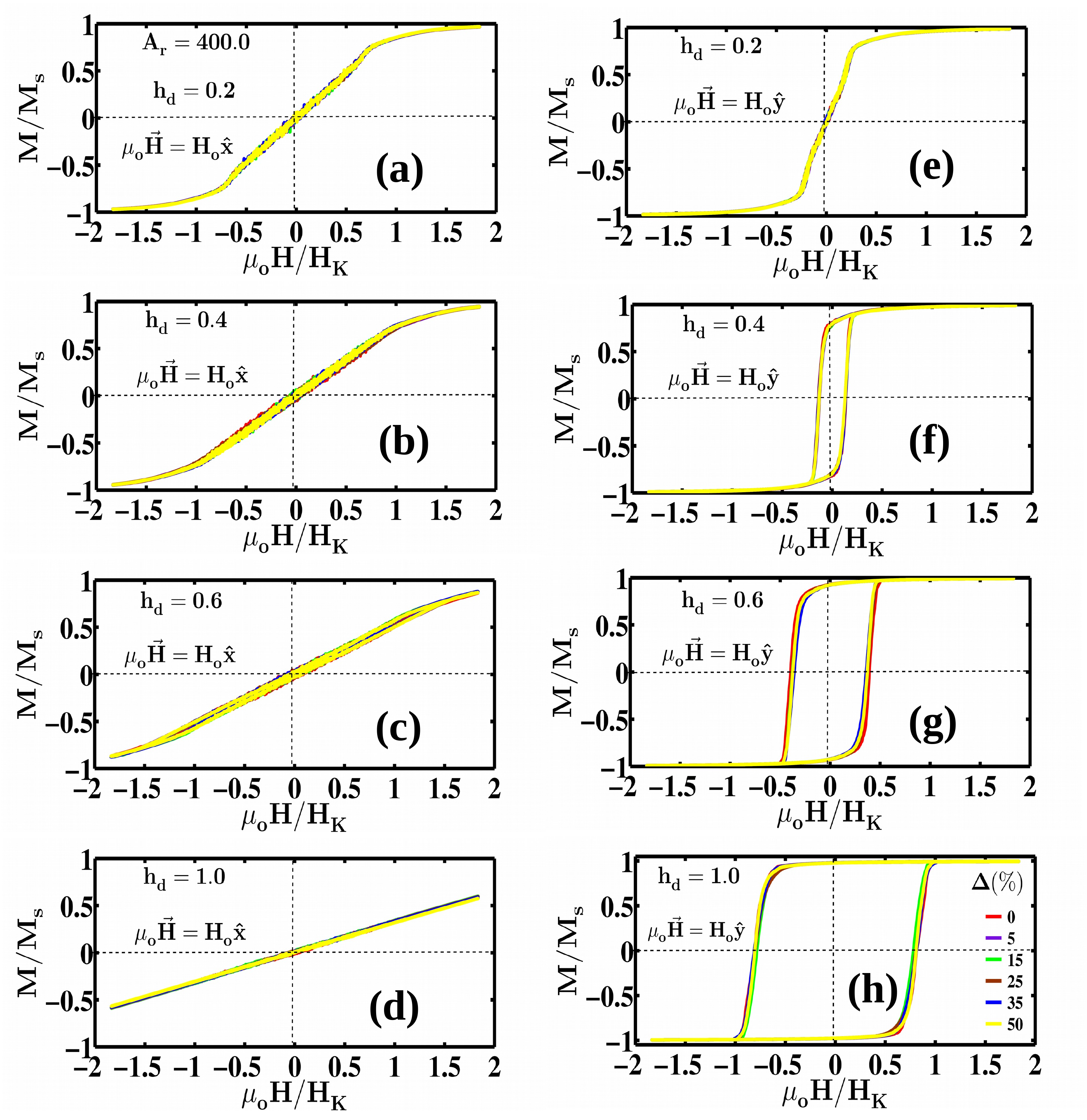}
\caption{Magnetic hysteresis in highly anisotropic system ($A^{}_r=400$), which corresponds to linear array of MNPs. The non-hysteresis is observed with the field along shorter axis of the system ($x$-axis). The coercive and remanence are very large with the field applied along long axis of the sample. It is because the dipolar interaction induces shape anisotropy along $y$-axis in such a case.}
\label{figure6}
\end{figure}

In Fig.~(\ref{figure6}), we study the hysteresis response with a highly anisotropic system $A^{}_r=400$, which corresponds to a one-dimensional chain of MNPs. MNPs stays in the superparamagnetic regime for small dipolar interaction strength even with huge $A^{}_r$, resulting in the minimal hysteresis loop area. We observed non-hysteresis with the field applied along the shorter axis of the sample ($x$-axis) with significant $h^{}_d$. These observations are independent of the disorder and dipolar interaction strength. Remarkably, the hysteresis loop area  enhances as $h^{}_d$ increases, provided the field is applied along the long axis of the system ($y$-axis). The dipolar interaction promotes ferromagnetic coupling in the highly anisotropic system, such as the linear and cylindrical arrangement of MNPs. Var{\'o}n {\it et al.} also found the similar behaviour of magnetic moments in low-dimensional system~\cite{varon2013}. Consequently, there is a natural tendency of magnetic moments to get aligned along the long axis of the sample even in the absence of an external field, resulting in a huge remanence($\sim 1.0$) and hysteresis loop area. The increase of hysteresis loop area with $h^{}_d$ is also in the perfect agreement with our recent work~\cite{anand2020}.  



\begin{figure}[!htb]
\centering\includegraphics[scale=0.50]{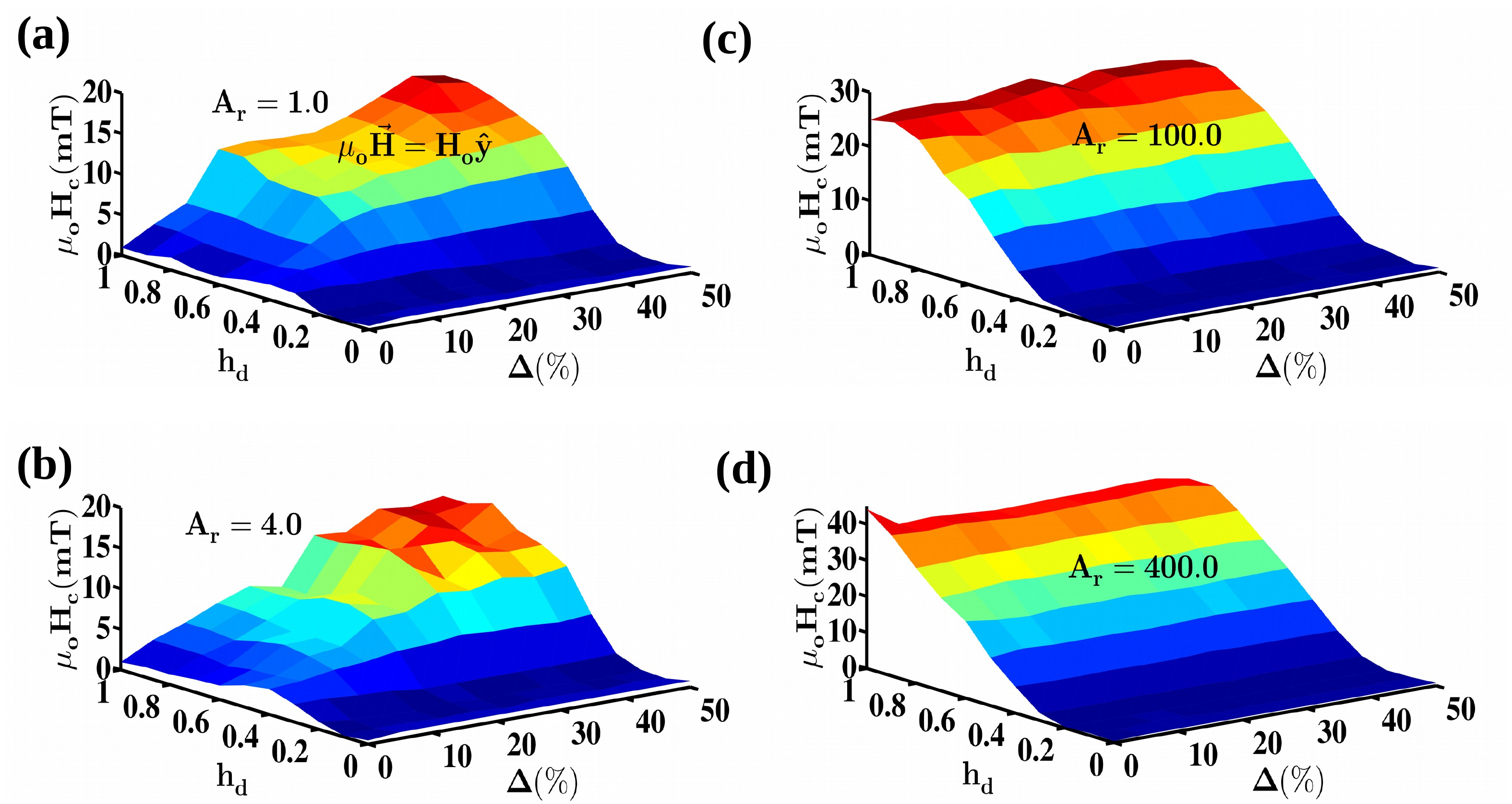}
\caption{The variation of the coercive field $\mu^{}_oH^{}_c$ as  a function of out-of-plane disorder $\Delta$ and interaction strength $\Delta$ with the external field along $y$-direction. We have considered four representative values of $A^{}_r=1.0$ [(a)], 4.0 [(b)], 100.0 [(c)] and 400.0 [(d)]. $\mu^{}_oH^{}_c$ is minimal for weakly interacting MNPs ($h^{}_d\leq0.2$), independent of $A^{}_r$ and $\Delta$, indicating superparamagnetic behaviour. In the presence of large $h^{}_d$, $\mu^{}_oH^{}_c$ increases with $\Delta$ in square-like MNPs arrays ($A^{}_r\leq4.0$). There is  always an increase in $\mu^{}_oH^{}_c$ with $h^{}_d$ for a fixed $\Delta$ because of an enhancement in ferromagnetic coupling.
} 
\label{figure7}
\end{figure}

\begin{figure}[!htb]
	\centering\includegraphics[scale=0.450]{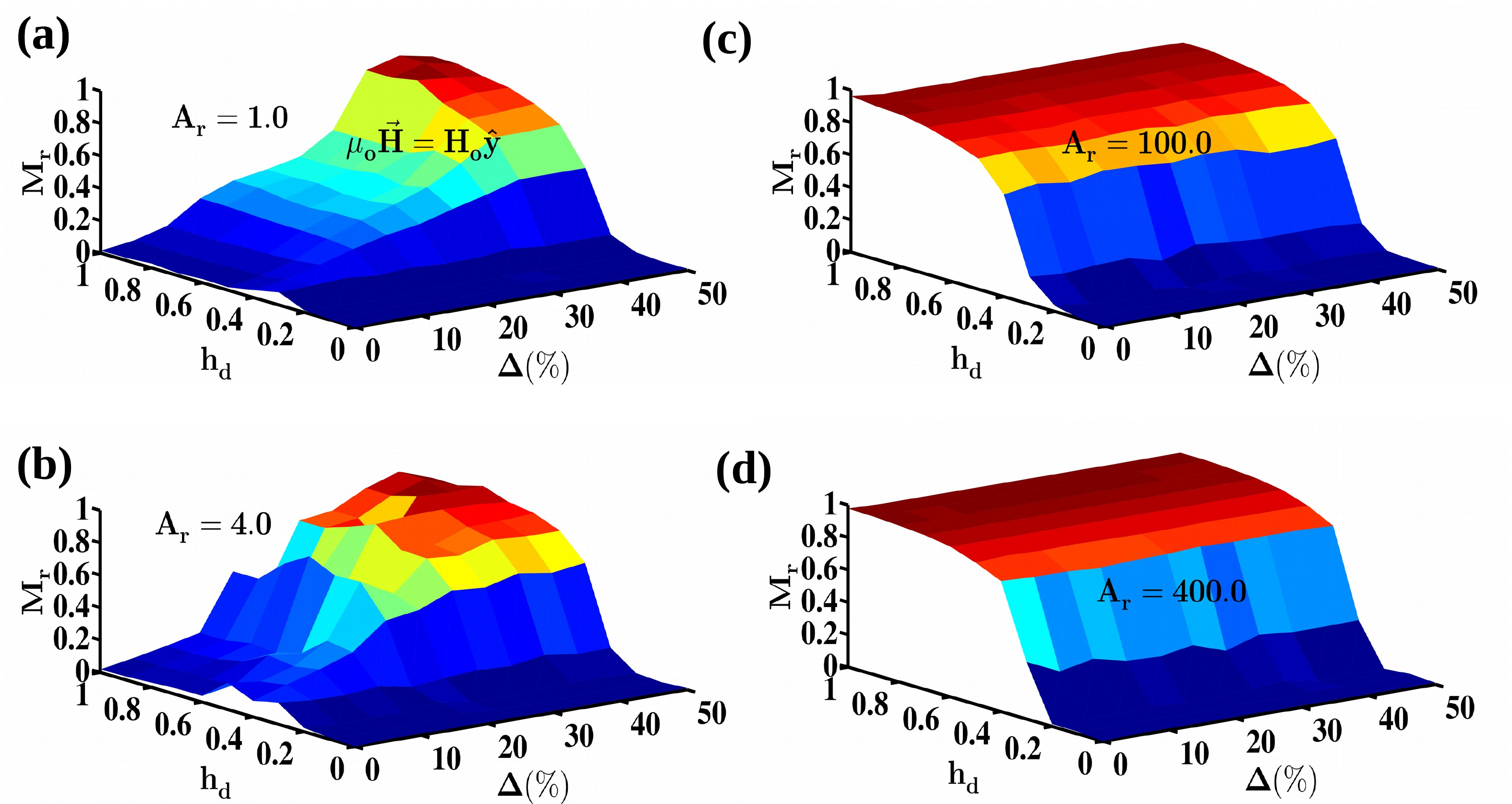}
	\caption{The variation of remanent magnetization $M^{}_r$ as a function of dipolar interaction strength $h^{}_d$ and out-of-plane disorder $\Delta$. We have considered four typical values of $A^{}_r=1.0$ [(a)], 4.0 [(b)], 100.0 [(c)] and 400.0 [(d)]. $M^{}_r$ increases with an increase in disorder strength $\Delta$ in the presenc of large $h^{}_d$ and smaller $A^{}_r$. $M^{}_r$ reaches to $\sim 1.0$ even with moderate $h^{}_d$ due to ferromagnetic coupling dominance.}
	\label{figure8}
\end{figure}

It is also essential to investigate the hysteresis mechanism quantitatively by analyzing the coercive field $\mu^{}_oH^{}_c$, remanence $M^{}_r$ and amount of heat dissipation $E^{}_H$ (hysteresis loop area) as a function of these parameters to have a better understanding of the hysteresis characteristics. Fig.~(\ref{figure7}) and Fig.~(\ref{figure8}) show the variation of $\mu^{}_oH^{}_c$ and $M^{}_r$ as a function of $\Delta(\%)$ and $h^{}_d$, respectively. We have considered four typical values of $A^{}_r=1.0$, 4.0, 100.0, and 400.0, and the external magnetic field is applied along the $y$-axis as exciting physics unfolds in such a case. $\mu^{}_oH^{}_c$ and $M^{}_r$ are negligibly small with weakly dipolar interacting MNPs ($h^{}_d\leq0.2$), irrespective of $\Delta$  and $A^{}_r$, indicating the superparamagnetic character. In the absence of disorder ($\Delta\approx0$), $\mu^{}_oH^{}_c$ and $M^{}_r$ have small values even with large $h^{}_d$ and more minor $A^{}_r$. It is because the dipolar interaction promotes antiferromagnetic coupling among the magnetic moments in the square-like arrangement of MNPs. On the other hand, there is an enhancement in $M^{}_r$ and $\mu^{}_oH^{}_c$ as dipolar interaction strength $h^{}_d$ increases for enormous $A^{}_r$ and fixed disorder strength. We can explain it by the fact the ferromagnetic interaction is promoted due to dipolar interaction in the highly anisotropic system (huge $A^{}_r$). Notably, the remanence $M^{}_r$ is huge $\sim 1.0$ even with moderate interaction strength ($h^{}_d\approx 0.6$) in a system with a large aspect ratio. The dipolar interaction induces shape anisotropy along the long axis of the system ($y$-axis in the present case) with huge $A^{}_r$, instigating magnetic moments to get aligned along $y$-direction even in the absence of an external magnetic field. The increase of $M^{}_r$ and $\mu^{}_oH^{}_c$ with $h^{}_d$ in the absence of $\Delta$ is perect agreement with the work of Figueiredo {\it et al.}~\cite{figueiredo2007}.  We could not compare our results in the presence of defects with them as they have focused only on the pefectly ordered MNPs ensembles.

\begin{figure}[!htb]
		\centering\includegraphics[scale=0.450]{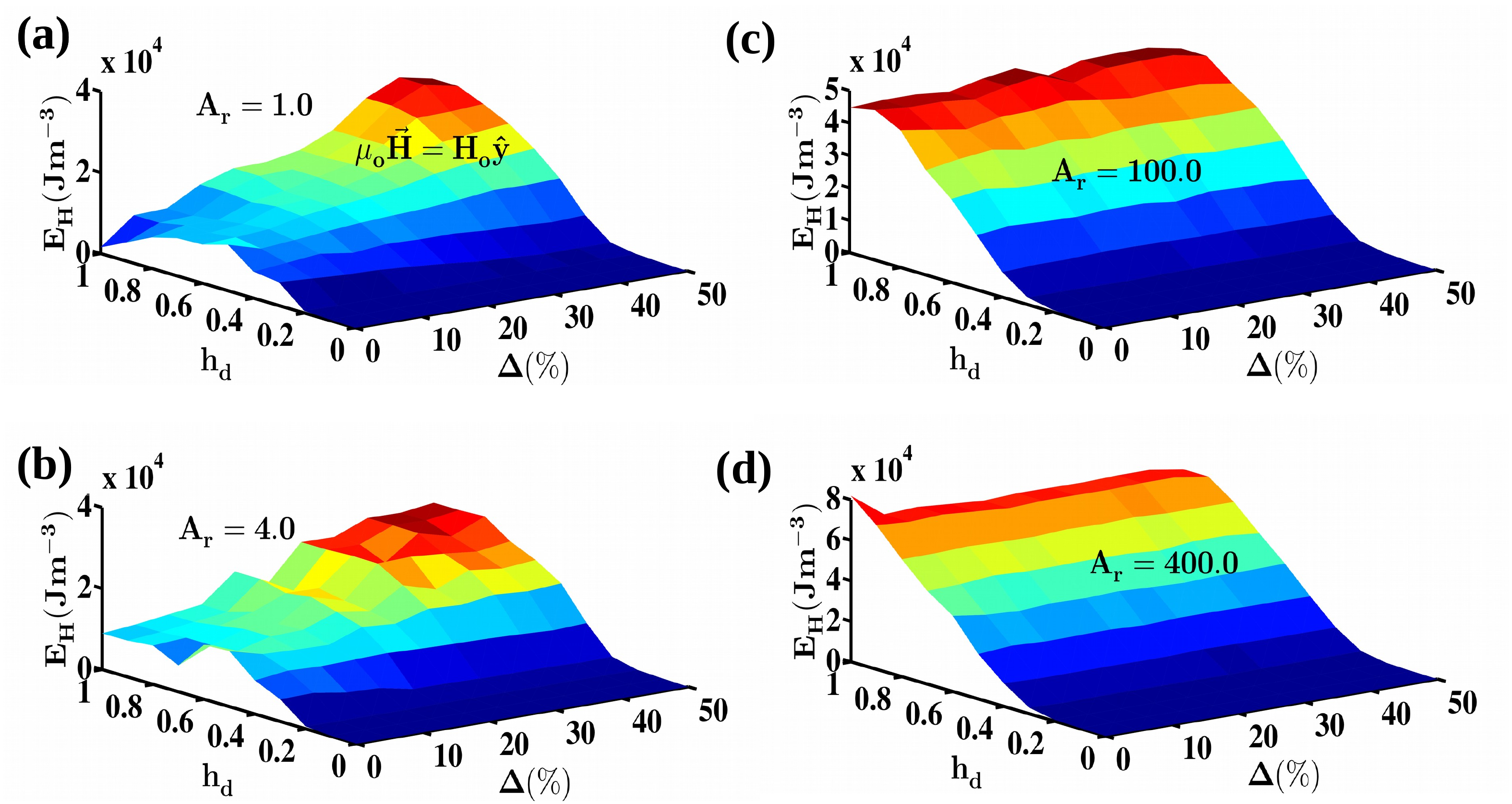}
		\caption{The variation of the amount of heat dissipated $E^{}_H$ due to the hysteresis as a function of out-of-plane disorder $\Delta$ and interaction strength $h^{}_d$. Four typical values of $A^{}_r=1.0$ [(a)], 4.0 [(b)], 100.0 [(c)], and 400.0 [(d)] are considered. The superparamagnetic character is dominant for weakly interating MNPs, resulting in minimal $E^{}_H$. There an enhancement in $E^{}_H$ with $\Delta$ for large $h^{}_d$ as the nature of interaction changes from antiferromagnetic to ferromagnetic. In the case of huge $A^{}_r$, dipolar interaction of suffiencent strength increases $E^{}_H$, independent of disorder strength.} 
		\label{figure9}
	\end{figure}

Finally, we study the variation of the heat dissipation $E^{}_H$ due to hysteresis as a function of disorder strength $\Delta$ and dipolar interaction strength $h^{}_d$ in Fig.~(\ref{figure9}). We have considered four representative values of aspect ratio $A^{}_r=1.0$, 4.0, 100.0, and 400.0. $E^{}_H$ has minimal value for weakly dipolar interacting MNPs, irrespective of $\Delta$ and $A^{}_r$, reminiscent of superparamagnetic behaviour. In a perfectly ordered state ($\Delta=0$), the antiferromagnetic coupling is dominant with significant $h^{}_d$ and relatively smaller $A^{}_r$. Consequently, the amount of heat dissipated $E^{}_H$ is tiny in these cases [see Fig.~\ref{figure9}(a)-(b)]. Remarkably, $E^{}_H$ increases with $\Delta$ in the presence of appreciable $h^{}_d$ and smaller $A^{}_r$. It is because the dipolar interaction of sufficient strength puts the system from antiferromagnetic to ferromagnetic state with an increase in out-of-plane disorder strength $\Delta$. On the other hand, $E^{}_H$ increases with $h^{}_d$ for a given $\Delta$ and huge $A^{}_r$ because of an enhancement in ferromagnetic coupling. Moreover, $E^{}_H$ depends weakly on out-of-plane disorder strength $\Delta$ in a such case as ferromagnetic interaction is already attained its maximum. 
	
\section{Summary and conclusion}
Two-dimensional arrays of MNPs are of immense importance due to their usages in various applications such as digital data storage, magnetic hyperthermia and spintronics, etc. Generally, the experimentally fabricated assays are not perfectly ordered~\cite{jensen2003scaling,xu2020two}.  In such cases, it is quite possible that few of the nanoparticles lie off-axis, normal to the sample plane~\cite{cheng2004magnetic}. 
Such positional defects are termed out-of-plane disorders. It is also a well-known fact that dipolar interaction is greatly affected by nanoparticles's interparticle separation and geometrical arrangement. Therefore, it is essential to investigate the magnetic properties as a function of out-of-plane disorder strength $\Delta$, dipolar interaction strength $h^{}_d$ and other system parameters such as aspect ratio $A^{}_r$ of the system.  In the present work, we analyzed the hysteresis characteristics in the two-dimensional arrays ($l^{}_x\times l^{}_y$) as a function of $\Delta$, $h^{}_d$, $A^{}_r=l^{}_y/l^{}_x$, and external magnetic field directions using kinetic Monte Carlo simulations. We have implemented the kinetic Monte Carlo simulation algorithm for these studies, which is superior to the Metropolis Monte Carlo technique for dynamical properties analysis. In the presence of weak dipolar interaction ($h^{}_d\leq0.2$), the hysteresis is dictated by superparamagnetic behaviour, resulting in the minimal hysteresis loop area. Our observation is in the perfect agreement with the work of Sousa {\it et al.}~\cite{de2005}. Remarkably, the dipolar interaction of sufficient strength promotes antiferromagnetic coupling in a square-like well ordered ($\Delta\approx0$) MNPs arrays. Consequently, the double hysteresis loop emerges, characteristics of antiferromagnetic coupling dominance. Exciting physics appears with the introduction of the out-of-plane disorder in the underlying system. Yang {\it et al.} also found the antiferromagnetic interaction dominance in nanoparticle arrays~\cite{yang2006roles}. Interestingly, the dipolar interaction of enough strength drives the system from antiferromagnetic to ferromagnetic magnetic state in the presence of large $\Delta$ even with smaller $A^{}_r$ (square assembled of MNPs). On the other hand, an additional anisotropy (shape anisotropy) is developed due to the dipolar interaction in the highly anisotropic system along the long axis of arrays ($y$-axis in the present case). As a consequence, the magnetic moments tend to align ferromagnetically, resulting in an enhancement in the hysteresis loop area with the field applied along the $y$-direction. The increase of area under hysteresis curve with interaction strength is in perfect qualitative agreement with the work of Kechrakos {\it et al.}~\cite{kechrakos2004}. In contrast, non-hysteresis is observed when the external field is applied shorter axis of the system, i.e. $x$-axis. These results suggest that we can tune the nature of dipolar interaction from antiferromagnetic to ferromagnetic by just inducing positional disorder even in the square assembly of MNPs. Therefore, these observations provide a concrete theoretical basis to manipulate the nature of dipolar interaction in a more controlled manner. We are also able to explain the observed hysteresis behaviour of such experimentally obtained MNPs assays.    

The quantitative analysis of coercive field $\mu^{}_oH^{}_c$, remanence $M^{}_r$ and the heat dissipation $E^{}_H$ variation with these parameters could help assess the hysteresis properties. In the presence of weak or negligible dipolar interaction, $\mu^{}_oH^{}_c$ and $M^{}_r$ are minimal irrespective of $A^{}_r$ and applied magnetic field directions, indicating superparamagnetic behaviour. Remarkably, $M^{}_r$ and $\mu^{}_oH^{}_c$ are extremely small with large $h^{}_d$ and smaller $A^{}_r$, quantifying the antiferromagnetic coupling dominance. Interestingly, there is an enhancement in these values as $\Delta$ is increased, provided $h^{}_d$ is large enough. The nature of dipolar interaction changes from antiferromagnetic to ferromagnetic in the presence of significant out-of-plane disorder strength $\Delta$ with relatively more minor $A^{}_r$. In the highly anisotropic system (huge $A^{}_r$), $\mu^{}_oH^{}_c$ and $M^{}_r$ have enormous values even in the presence of moderate $h^{}_d$ with  the field applied along the long axis of the system. In contrast, they are minimal when the field is along the shorter axis of the sample, indicating non-hysteretic behaviour. Notably, $E^{}_H$ varies with these parameters exactly same manner as that of the coercive field, implying an intimate relationship between them.

In conclusion, we analyse the hysteresis properties in the two-dimensional arrays of MNPs with the out-of-plane disorder to provide a theoretical basis for unusual characteristics in such systems. The role of dipolar interaction is strongly affected by the disorder strength, aspect ratio of the system. In particular, the dipolar interaction of enough strength drives the magnetic order from antiferromagnetic to ferromagnetic in the square-like arrangement of MNPs. While in the highly anisotropic system, even the moderate dipolar interaction is able to induce ferromagnetic coupling among the magnetic moments. The results obtained in the present work should help the experimentalist to explain the physics of the observed hysteresis response for the nanofabricated MNPs assays. These are also useful in manipulating the nature of dipolar interaction in a more controlled manner analytically.  
Therefore, we hope that the present work could pave the way for joint efforts in experimental, theoretical, and computational studies for these excellent and versatile systems.


\section*{DATA AVAILABILITY}
The data that support the findings of this study are available from the corresponding author upon reasonable request.
\newpage
\bibliography{ref}
\newpage

\newpage
\newpage

\newpage

\newpage

\newpage
\newpage
\newpage
	
	\newpage
\end{document}